\documentclass[twocolumn,showpacs,preprintnumbers, 
amsmath,amssymb,aps,prd,nofootinbib,superscriptaddress, 
eqsecnum,floatfix]{revtex4} 
 
\def\sl#1{\setbox0=\hbox{$#1$}               
   \dimen0=\wd0                                 
   \setbox1=\hbox{/} \dimen1=\wd1               
   \ifdim\dimen0>\dimen1                        
      \rlap{\hbox to \dimen0{\hfil/\hfil}}      
      #1                                        
   \else                                        
      \rlap{\hbox to \dimen1{\hfil$#1$\hfil}}   
      /                                         
   \fi}                                         %
\usepackage{graphicx,color,amsmath} 
 
\makeatletter 
\renewcommand{\p@subsection}{} 
\makeatother

\def\Eq#1{Eq.~(\ref{#1})} 
\def\Fig#1{Fig.~\ref{#1}} 
\def\Ref#1{Ref.~\cite{#1}} 
 
\newcommand{\vev}[1]{\langle{#1}\rangle} 
\newcommand{\frgg}{FRG$_{\rm grid}$} 
\newcommand{\frgt}{FRG$_{\rm Taylor}$} 
 
\newcommand\nn{\nonumber \\}

\def\sla{\slash{\!\!\!}}

\newcommand{\lk}{\left(} 
\newcommand{\rk}{\right)}

\newcommand{\ldk}{\left[} 
\newcommand{\rdk}{\right]} 
\newcommand\beq{ \begin{eqnarray} } 
\newcommand\eeq{ \end{eqnarray} } 

\graphicspath{ 
{./} 
{../figures/} 
} 
\begin{document} 
 
\title{Fluctuations and isentropes near the chiral
  critical endpoint}

\author{E.~Nakano}
\email[E-mail:]{e.nakano@gsi.de} 
\affiliation{Extreme Matter Institute EMMI, GSI, Planckstr.~1,  D-64291
  Darmstadt, Germany} 
 
\author{B.-J.~Schaefer} 
\email[E-mail:]{bernd-jochen.schaefer@uni-graz.at} 
\affiliation{Institut f\"{u}r Physik, Karl-Franzens-Universit\"{a}t, 
  A-8010 Graz, Austria} 
 
\author{B.~Stokic} 
\email[E-mail:]{b.stokic@gsi.de} 
\affiliation{GSI Helmholtzzentrum f\"ur Schwerionenforschung, Planckstr.~1,
D-64291
  Darmstadt, Germany} 

\author{B.~Friman} 
\email[E-mail:]{b.friman@gsi.de} 
\affiliation{GSI Helmholtzzentrum f\"ur Schwerionenforschung, Planckstr.~1,
D-64291
  Darmstadt, Germany} 
 
\author{K.~Redlich} 
\email[E-mail:]{redlich@ift.uni.wroc.pl} 
\affiliation{Institute of Theoretical Physics, University of Wroclaw, 
PL-50204 Wroclaw, Poland}

 
\date{\today} 
 
\begin{abstract} 
  Isentropic trajectories crossing the chiral phase transition near
  the critical endpoint (CEP) are studied for two light quark flavors.
  The calculations are performed within an effective chiral model with
  quark-meson interactions, belonging to the same universality class
  as QCD. We confront mean-field thermodynamics with the functional
  renormalization group approach, where fluctuations are properly
  taken into account. We establish a connection between modifications
  of the isentropic trajectories found in mean-field calculations at
  the crossover transition near the CEP and the order of the phase
  transition in the chiral limit. Furthermore, the isentropes obtained
  with the renormalization group are completely smooth at the
  crossover transition and do not in any way reflect the proximity of
  the CEP. In particular, our results do not show the recently
  conjectured focusing of isentropes from the crossover region
  towards the critical endpoint.
\end{abstract}
\maketitle


\section{Introduction}

The determination of the phase structure and the critical behavior of
strongly interacting matter is one of the major goals addressed
theoretically in studies of QCD at finite temperature $T$ and quark
chemical potential $\mu$ and experimentally in ultrarelativistic
nuclear collisions. One of the fundamental predictions of QCD, is the
existence of a boundary line in the $(T,\mu)$-plane that separates the
confined, chirally broken hadronic phase from the deconfined chirally
symmetric quark-gluon plasma (QGP) \cite{Kogut:2004su}.

The existence of such a phase boundary was recently established by
first-principle calculations in Lattice Gauge Theory (LGT)
\cite{finite-mu}. For vanishing $\mu$ and for two massless
flavors, the chiral transition in the presence of an axial anomaly was argued
\cite{pisarski} to be second-order and in the universality class of
the $O(4)$-spin model. For finite quark masses, due to the explicit
breaking of the chiral symmetry, the second-order transition is most
likely replaced by a rapid crossover. Moreover, arguments based on
effective models \cite{stephanov:2004wx, hatta, Mocsy, PNJL,
  Fujii:2003bz, Sasaki:2006ws, Sasaki:2006ww, Stokic:2008jh,
  Schaefer:2007pw, Schaefer:2008hk, Schaefer:2006ds} indicate that at
large $\mu$ and moderate temperatures the transition along the phase
boundary is first-order.

The different nature of the transition at high and low $\mu$ suggests
that the QCD phase diagram exhibits at least one critical endpoint
(CEP), where the first-order phase transition line ends in a
second-order critical point, followed by a crossover region
\cite{stephanov:2004wx, Yagi:2005yb, Gebhardt1980, Berdnikov:1999ph}.
It is expected that the static critical properties of the second-order
chiral endpoint are governed by the universality class of the 3D Ising
model \cite{ising}. The universal properties of the phase diagram have
recently been studied extensively in an effective theory within the
functional renormalization group (FRG) approach,
e.g.~\cite{Berges:1997eu, Schaefer:2006ds, Schaefer1999, SFR}.
 
Although there is still no direct proof for the existence of the CEP in
the QCD phase diagram, many phenomenological models predict such a
point. However, its location is still strongly
model dependent \cite{stephanov:2004wx}. Lattice QCD simulations
also provide indications for a possible CEP in the QCD phase
diagram~\cite{finite-mu}.

The critical behavior and the position of the CEP can be identified by
means of observables sensitive to the singular part of the free
energy. Of particular interest in this context are observables which
reflect the fluctuations of conserved charges \cite{hatta,
 Fujii:2003bz, Sasaki:2006ws, Sasaki:2006ww, Schaefer:2006ds,
 shuryak,stephanovh}. For systems in thermodynamic equilibrium, both baryon
number and charge density fluctuations diverge at the CEP
\cite{stephanov:2004wx}, independently of the value of the quark mass.
However, a similar divergence is expected also at a first-order phase
transition in non--equilibrium situations, where the spinodal
instabilities are reached \cite{spinodal}.

The hydrodynamic expansion of an ideal fluid follows trajectories of constant
entropy, the so called isentropes. Due to baryon-number
conservation, such trajectories correspond to contours of constant
entropy per baryon $s/n$ in the temperature--chemical potential plane.
In Ref.~\cite{Stephanov:1998dy} it was pointed out that an expanding
system, which follows an isentropic trajectory crossing a first-order
phase transition between the quark-gluon plasma and the hadronic
phase, is focused towards the CEP, if $s/n$, at a
given point $(\mu,T)$ on the phase boundary, is larger in the
quark-gluon plasma. This implies that a larger range of initial
conditions will end up in the vicinity of the CEP.

Recently it was argued that the CEP acts as an attractor for
isentropic trajectories also on the crossover side, i.e., for values
of the chemical potential $\mu$ smaller than the value at the
endpoint~\cite{Nonaka:2004pg, Bluhm:2006av}. This feature could
potentially be used to experimentally verify the existence of the CEP
in ultrarelativistic nucleus-nucleus collisions~\cite{Asakawa:2008ti,
  Luo:2009sx}.

We note that the relevance of the isentropic trajectories for
relativistic nucleus-nucleus collisions rest on the assumption that
ideal hydrodynamics provides a good approximation to the true
expansion dynamics and hence that dissipative effects can be
neglected. Indeed, recent analyses of relativistic heavy-ion collider
(RHIC) data, show that the QGP, created in nucleus-nucleus collisions,
behaves as an almost ideal fluid with a very small $\eta/s$, the ratio
of shear viscosity to entropy density \cite{Teaney:2000cw}
This implies, that
dissipative processes in the QGP are strongly suppressed, leading to
an essentially isentropic expansion towards the phase boundary
\cite{Rischke:1995pe}.

However, this is not the case at the CEP~\cite{onuki} where dynamical
scaling implies that both, the shear and bulk viscosities,
diverge.~\footnote{An increase of the bulk viscosity near the CEP
\cite{Kharzeev:2007wb} is suggested by model calculations \cite{cs}
and LGT studies \cite{Meyer:2007dy}.} Hence, close to the CEP the
expansion is most likely not isentropic and consequently the relevance
of isentropic trajectories is questionable. Closely related to this
issue is the critical slowing down of long-wavelength fluctuations
close to a second-order phase transition, which implies that the
equilibration time of such a system diverges at the critical endpoint.
It follows that as the CEP is approached, it becomes increasingly
unlikely that an expanding system remains in equilibrium, i.e.,
expands isentropically. However, let us for the moment ignore this
problem, and assume that the expansion of the system is sufficiently
slow, so that local thermodynamic equilibrium is maintained
everywhere.

Under this assumption, the trajectory of the fireball produced in a
nucleus-nucleus collision follows an isentropic trajectory. The
corresponding value of $s/n$ is a function of the collision energy.
Thus, in such a system, the isentropes in the $(T,\mu)$-plane encode
important information on the expansion dynamics. Consequently,
possible modifications of the isentropes near the CEP influences the
evolution of the fireball in a characteristic way, which could provide
an experimental signature for the existence of the endpoint. The
search for the CEP is one of the objectives of planned nucleus-nucleus
collision experiments at RHIC/BNL~\cite{Adcox:2004mh} and of future
experiments at FAIR/GSI~\cite{Hohne:2006fr}.
 
In this work we investigate the properties of isentropic trajectories
near the CEP within an effective chiral theory. We examine the model
dependence of the focusing on the crossover side of the CEP
\cite{Nonaka:2004pg} both within the mean-field approach and in a
functional renormalization group (FRG) analysis, where fluctuations
and nonperturbative effects are properly accounted for. The FRG method
has been successfully employed to describe a broad range of critical
phenomena~\cite{Schaefer:2006ds, Berges:1997eu,Schaefer1999, SFR, Braun:2007td}.
Consequently, this method is ideally suited for exploring the effect
of fluctuations on the isentropic trajectories. 
Note that, due to the lack of gluonic degrees of freedom in the model,
the entropy density is much below that of QCD matter at the same temperature 
and chemical potential. Nevertheless, since the model 
is most likely in the universality class of two flavor QCD,
it can provide guidance on universal properties like e.g. 
critical exponents and the conjectured focusing of the isentropes towards the
CEP.
 
The paper is organized as follows. In the next section we introduce
the effective quark-meson model and the FRG approach. In
Sec.~\ref{sec:mfa} we discuss the mean-field thermodynamics, while in
Sec.~\ref{sec:isentropic} we present the properties of the isentropic
trajectories near the chiral phase transition. Finally, we summarize
our results in Sec.~\ref{sec:summary}.

\section{Flow equation for a chiral effective theory} 
 
The quark-meson model is a low-energy effective theory, which
incorporates the chiral symmetry of QCD. For two quark flavors and
$SU(3)_c$ color symmetry the model Lagrangian reads
\begin{equation} 
  \mathcal{L}= 
  \bar{\psi} \left[ i \sla\partial 
    -g \lk \sigma+ i\gamma_5 \vec{\tau} \cdot \vec{\pi} \rk \right] \psi
  +\frac{1}{2}\lk \partial_\mu  \phi \rk^2 -U(\sigma, \vec{\pi}), 
\label{eq:qmmodel} 
\end{equation} 
where $\phi=\lk \sigma, \vec{\pi} \rk$ is the $O(4)$-representation
for the isoscalar $\sigma$- and the isovector $\vec\pi$-mesons. The
two-flavor quark field $\psi$ couples to mesons via the flavor-blind
Yukawa coupling constant $g$. The purely mesonic potential is
given by
\begin{equation}\label{eq:potential} 
  U(\sigma, \vec{\pi})=\frac{1}{2} m^2 \phi^2 + 
  \frac{\lambda}{4} \lk \phi^2\rk ^2 - h\sigma\ , 
\end{equation} 
where $\lambda>0$ is the mesonic self-coupling. For a negative $m^2$,
the vacuum exhibits spontaneous chiral symmetry breaking, where the
chiral $SU(2)_L \times SU(2)_R$ symmetry of the Lagrangian is broken
down to the $SU(2)$ vector symmetry. At high temperatures and/or
densities the spontaneously broken chiral symmetry is restored. The
external field $h$, which is related to the current quark masses,
breaks the chiral symmetry explicitly.
 
In the chiral quark-meson model, explicit gluonic degrees of freedom
are missing. The effect of gluons is to a certain extent implicitly
included in the coupling constants. However, recent extensions of
  the quark-meson model in this context (e.g.~by including the Polyakov loop)
  are possible \cite{Schaefer:2007pw} and currently under
  investigation.

\subsection{FRG equation for the effective potential} 
 
In this section we present the flow equation for the thermodynamic potential
$\Omega(T,\mu)$ in the FRG approach. Within this scheme, we then
compute the thermodynamic potential as well as the entropy and baryon
number densities near the chiral phase transition, including the
effects of fluctuations. The FRG method yields the average effective
action $\Gamma_k$ at the momentum scale $k$. This scale is introduced
in the flow equation via regulator functions, which act as mass terms
in the propagators. The effect of the regulators is to suppress the
propagation of particles with momenta smaller than $k$. In the
infrared limit ($k=0$) fluctuations at all wavelengths have been
integrated out and $\Gamma_{k=0}$ is the full effective action.

The FRG flow equation smoothly interpolates the physics between the
ultraviolet (UV) $\Lambda$ and the IR scale \cite{Wetterich:1992yh,
  Berges:2000ew, Schaefer:2006sr}. For the quark-meson model it
contains two terms $(t = \ln (k/\Lambda) )$
\begin{eqnarray} 
  \partial_t \Gamma_k[\Phi, \Psi] \!\!&\!=\!&\!\!\frac{1}{2}{\rm Tr} \ldk \partial_t 
  R_{B, k}\lk \Gamma_{B, k}^{(2)}[\Phi, \Psi] +R_{B, k} \rk^{-1} \rdk 
  \nonumber \\ 
&&\!\!\!\!\!-{\rm Tr} \ldk \partial_t R_{F, k}\lk \Gamma_{F, k}^{(2)}[\Phi, 
  \Psi] +R_{F, k} \rk^{-1} \right]\!.
  \label{eq:frg} 
\end{eqnarray} 
The first term in \Eq{eq:frg} represents the bosonic flow with the
regulator $R_{B, k}$ while the second part stands for the fermionic
contribution with the regulator $R_{F, k}$. The bosonic and fermionic
fields are denoted by $\Phi$ and $\Psi$, respectively. The full
bosonic (fermionic) inverse propagator includes the term
$\Gamma_{B (F), k}^{(2)}$, the second functional derivative of
$\Gamma_{k}$ with respect to the corresponding fields $\Phi$ or
$\Psi$.

As initial condition for the flow equation (\ref{eq:frg}) one chooses
a bare effective action at the UV scale $\Lambda$. Consequently, at
the momentum scale $\Lambda$, $\Gamma_{k}$ coincides with the
classical action $S$, i.e., $\Gamma_{k=\Lambda}\equiv S$. During the
evolution, the initial bare action is renormalized and finally, at
$k=0$, corresponds to the full effective action $\Gamma_{k=0}=
\Gamma$.

\subsection{Leading order derivative expansion} 

The functional flow equation (\ref{eq:frg}) is exact and is equivalent
to an infinite tower of coupled partial differential equations for
$n$-point functions ($n\ge2$). In order to solve this equation, a
suitable approximation scheme is required.

In the leading order (LO) derivative expansion the FRG flow equation
reduces to an ordinary non-linear differential
equation~\cite{Jungnickel:1995fp, Nicoll:1974zz}. This can be solved
either directly, for the effective potential on a momentum-space grid
or by expanding it in powers of the order parameter around a minimum.
In this letter we use both methods and compare the results. For the
quark-meson model, in four Euclidean space-time dimensions, the LO
derivative expansion yields~\cite{Jungnickel:1995fp, Berges:1997eu}
\begin{eqnarray} 
  \Gamma_k [\psi,\bar\psi,\phi]&=&\int {\rm d}^4x \ldk \bar{\psi}
  \lk \sla{\partial} +g \lk \sigma +i \vec{\tau} \cdot \vec{\pi}
  \gamma_5 \rk \rk \psi \right. \nonumber \\
&& \qquad \qquad \left. + \frac{1}{2} \lk \partial_\mu \phi \rk^2 
  +U_k(\rho)\rdk,
\end{eqnarray} 
where $U_k(\rho)$ is the scale-dependent effective
potential, cf.~\Eq{eq:qmmodel}. In this expression we have introduced
a symmetric field variable $\rho$ defined by $\rho=\frac{1}{2}
\phi^2=\frac{1}{2} \lk \sigma^2+\vec{\pi}^2 \rk$. For a uniform system
it is convenient to deal with the thermodynamic potential
density $\Omega_k = (T/V) \Gamma_{k}= U_k(\rho=\rho_{0,k})$,
where the scale-dependent minimum of the potential is labelled by $\rho_{0,k}$.
In this exploratory work, we neglect the scale evolution of the Yukawa
coupling $g$. This approximation does not significantly affect the
critical properties~\cite{Jungnickel:1995fp,Berges:1997eu,Braun:2008pi}.
 
In thermal equilibrium the extension of the flow equation to finite
$T$ and $\mu$ is done within the Matsubara formalism: the integration
over $p_0$ is replaced by a summation over the corresponding Matsubara
frequencies: $p_0 \rightarrow 2n\pi T $ for bosons and
$p_0 \rightarrow (2n+1)\pi T$ for fermions.
A finite chemical potential is introduced in the fermionic part of the
Lagrangian by shifting the derivative with respect to the Euclidean
time $\tau$, $\partial_\tau \rightarrow \partial_\tau-\mu$. We assume
$SU_f(2)$-symmetry and set $\mu=\mu_u=\mu_d$.

We employ the optimized regulators \cite{Litim:2000ci,Blaizot:2006rj}
for fermions
\begin{eqnarray} 
 \!\!\!\!\!\! R_{F, k}(q)\!\!&=&\!\!\lk \sla{q}+i\mu \gamma_0\rk \lk
  \sqrt{\frac{\tilde{q}_0^2+k^2}{\tilde{q}_0^2+\mathbf{q}^2}}-1 \rk\! \theta
\!\lk
  k^2-\mathbf{q}^2 \rk 
\end{eqnarray} 
and bosons 
\begin{eqnarray} 
  R_{B, k}(\mathbf{q})&=&\lk k^2-\mathbf{q}^2 \rk \theta\lk k^2-\mathbf{q}^2
  \rk, 
\end{eqnarray} 
where $\tilde{q}_0=q_0+i\mu$. With these, the
momentum integration and Matsubara summation in the RG flow equation can be done
analytically, resulting in the following flow equation for the thermodynamic
potential density $\Omega_k$
\begin{eqnarray} 
  &&\!\!\!\!\!\!\!\!\partial_k\Omega_k\lk T,\mu; \rho_{0,k} \rk 
  =\,\frac{k^4}{12 \pi^2}\!\!
  \ldk 3\frac{1+2n_B(E_\pi)}{E_\pi} \right. \nn 
&&\!\!\!\!\!\left.
  +\frac{1+2n_B(E_\sigma) }{E_\sigma} -2
\nu_q\frac{1-n^+_F(E_q)-n^-_F(E_q)}{E_q} \rdk.
\label{floweq0} 
\end{eqnarray} 
In \Eq{floweq0} $n_B(x)=\ldk
e^{x/T}-1\rdk^{-1}$ denotes the Bose-Einstein and $n^\pm_F(x)=\ldk
e^{(x\mp\mu)/T}+1\rdk^{-1}$ the Fermi-Dirac distribution functions for
bosons, quarks and antiquarks, respectively. The
single-particle energies of pions, sigmas and quarks are given by
$E_{\pi,\sigma,q}=\sqrt{k^2+M^2_{\pi,\sigma,q}}$ where the masses are scale
dependent:
$M^2_\pi=\bar{\Omega}_k'$, $M^2_\sigma=\bar{\Omega}_k'+2\rho_{0,k}
\bar{\Omega}_k''$ and
$M_q^2=2g^2 \rho_{0,k}$.  The prime on the potential
denotes the derivative of $\bar{\Omega}_k\,=\,\Omega_k\,+\,h\,\sqrt{2\,\rho}$
with respect to $\rho$ evaluated
at the minimum $\rho_{0,k}$. The quark degeneracy factor is
$\nu_q=2 N_f N_c=12$.

Although the RG flow equation (\ref{floweq0}) looks rather innocuous, 
it is quite powerful due to non-linearity implied by the self-consistent
determination of the single-particle energies. This scheme very efficiently
accounts for long-range fluctuations and nonperturbative dynamics near the
chiral phase transition~\cite{Litim:2000ci, Schaefer:2006sr}.

\subsection{Solving the FRG flow equation}
\label{sec:frg} 
 
In order to solve the flow equation (\ref{floweq0}), we employ two
distinct methods. The first one is the grid method, where the
potential is discretized on a one-dimensional $\rho_i$ grid. This
leads to a set of coupled flow equations for the scale-dependent
potential $\Omega_k (T,\mu; \rho_i)$ at each grid point $\rho_i$.
Using the bare potential as initial condition at a given UV cutoff
$\Lambda$, the scale evolution of the thermodynamic potential is
obtained by finding the minimum with respect to variations of the
field $\rho$ at each scale. For details concerning the numerical
implementation we refer to Ref. \cite{Bohr2001}.

The second method is based on a Taylor-expansion, up to a maximum power $N$, of
$\bar{\Omega}_k$ around the scale-dependent running minimum $\rho_{0,k}
=\sigma_{0,k}^2/2$
\begin{equation} 
  \bar{\Omega}_k(T,\mu; \rho)=\sum_{m=0}^N \frac{a_{m,k}}{m!}
  (\rho-\rho_{0,k})^m 
  \label{eq:taylor} . 
\end{equation} 
The coefficients $a_{m,k}$ are functions of the scale $k$ and of
the temperature and chemical potential.
 
On the one hand, the expansion method yields the potential only in a
limited range around the minimum, due to the finite convergence radius of the
Taylor expansion. It is therefore difficult to describe a first-order phase
transition, where two local potential minima have to be considered. On the other
hand, the advantage of the Taylor expansion is its simplicity; only $N+1$
coupled differential equations have to be solved.
 
The minimum of the thermodynamic potential is determined by the stationarity
condition
\begin{equation} 
  \left.\frac{{\rm d} \Omega_k}{{\rm d} 
      \sigma}\right|_{\sigma=\sigma_{0,k}}=\left.\frac{{\rm d}
\bar{\Omega}_k}{{\rm d}
      \sigma}\right|_{\sigma=\sigma_{0,k}}-h=0.
\label{gapeq} 
\end{equation} 
In the Taylor expansion scheme this yields a relation between the coupling $a_1$
and the expectation value of
the scalar field $\sigma_0$
\begin{equation} 
  a_{1,k} = h/ \sigma_{0,k}\ . 
\label{gapeq1} 
\end{equation} 
We truncate the
expansion up to the third order, $N=3$, and obtain the following set
of flow equations:
\begin{eqnarray} 
  \partial_k a_{0,k} 
  &=& \frac{h}{\sqrt{2\rho_{0,k}}}\,\partial_k\, \rho_{0,k} + \partial_k \,
  \Omega_k\ , 
  \label{flowset01} \\ 
  \partial_k \rho_{0,k} 
  &=&-\,\frac{\partial_k \Omega'_k}{ h/(2\rho_{0,k})^{3/2}+a_{2,k}}\ , 
  \label{flowset02}\\ 
  \partial_k a_{2,k} 
  &=& a_{3,k} \,\partial_k \,\rho_{0,k} + \partial_k \,\Omega''_k\ , 
  \label{flowset03}\\ 
  \partial_k a_{3,k} 
  &=&\partial_k\, \Omega'''_k\ . 
  \label{flowset04} 
\end{eqnarray} 
The meson masses are then given by
\begin{equation} 
  M_{\pi, k}^2
  =\frac{h}{\sqrt{2\rho_{0,k}}}, \quad M_{\sigma, 
    k}^2 
  =\frac{h}{\sqrt{2\rho_{0,k}}}+2 \,\rho_{0,k}\, a_{2,k}, 
\label{masses01} 
\end{equation} 
and the quark mass by $M_{q, k}^2 =2\, g^2\, \rho_{0,k}$.
 
The flow equations (\ref{flowset01})-(\ref{flowset04}) are solved
numerically starting from the initial conditions for the coefficients
$a_{i,\Lambda}$ ($i=0,1,2,3$) at the ultraviolet scale $\Lambda$. The
model parameters of the Lagrangian (\ref{eq:qmmodel}) are
chosen such that at the scale $k=0$, the chiral symmetry is spontaneously broken
in vacuum with a pion mass of $M_\pi=138$ MeV and a finite expectation value of
the sigma field $\langle \sigma \rangle=\sigma_0$, which is identified with the
pion decay constant $f_\pi=93$ MeV. This, together with our choice for the sigma
mass in vacuum $M_\sigma=670$ MeV and the constituent quark mass $M_q=335$ MeV,
are obtained with $h = 1.771 \times 10^6$ MeV$^3$, $g = 3.6$, $\Lambda =950$
MeV, $a_{1,\Lambda}=(582 \mbox{MeV})^2$, $a_{2,\Lambda}=35.2$ and
$a_{3,\Lambda}=0$. The corresponding values for the parameters of the Lagrangian
(\ref{eq:qmmodel}) are $\lambda=a_{2,\Lambda}/2$
and $m^2=a_{1,\Lambda}-\lambda\,\sigma_{0,\Lambda} ^2$, where
$\sigma_{0,\Lambda}=h/a_{1,\Lambda}\simeq 5.2$ MeV is the starting value for the
scalar condensate.
 
\subsection{Isentropic thermodynamics}

The solution of the flow equation (\ref{floweq0}) yields the
pressure $p(T,\mu)=-\left.\Omega_{k}(T,\mu; \rho_{0,k}) \right|_{k=0}$ as a
function of $T$ and $\mu$.
The relevant  observables,  the entropy density $s(T, 
\mu)$ and the quark-number density $n(T,\mu)$ are obtained from the pressure as
derivatives with respect to temperature
and quark chemical potential
\begin{eqnarray} 
  s&=&\frac{ \partial p(T,\mu)}{\partial T}
  =-\left.\frac{\partial \ldk a_{0,k} - h\sigma_{0,k} \rdk} {\partial 
      T}\right|_{k=0}\ , \\ 
  n&=&\frac{ \partial p(T,\mu)}{\partial \mu}
  =-\left.\frac{\partial \ldk a_{0,k} - h\sigma_{0,k} \rdk} {\partial 
      \mu}\right|_{k=0}\ . 
\end{eqnarray} 
The isentropic trajectories in the $(T,\mu)$-plane are then obtained as contours
of constant entropy per baryon $s/n$.

\section{Mean-field approximation} 
\label{sec:mfa}
 
In order to examine the influence of fluctuations on the
thermodynamics, in particular on the isentropic trajectories, we
compare the FRG results with those obtained in the mean-field (MF)
approximation, where quantum and thermal fluctuations are neglected.
 
The partition function of the quark-meson model can be formulated as a
path integral over meson and quark/antiquark fields in Euclidean
space-time. In the MF approximation, the meson fields in the action
are replaced by their expectation values. The integration over the
quark/antiquark fields yields the fermionic determinant. The resulting
thermodynamic potential of the chiral quark--meson model is of the
form \cite{Scavenius:2000qd,Schaefer:2006ds}
\begin{equation}\label{pot} 
  \Omega(T,\mu; \vev\sigma , \vev{\vec{\pi}})  = \Omega_{\bar 
    qq}(T,\mu; \vev \sigma)  + U(\vev \sigma, \vev{\vec{\pi}}) 
\end{equation} 
with the meson potential $U$ of \Eq{eq:potential}. The
quark/anti\-quark contribution is given by 
\begin{eqnarray}\label{quark} 
\Omega_{\bar qq}(T,\mu; \vev \sigma) &=&  \nu_q T\int\!\frac{ 
  d^3k}{(2\pi)^3} \left\{ \ln\left(1-n_F^-(E_q)\right)\right.\nonumber  \\ 
&& \quad \left. + \ln\left(1-n_F^+(E_q)\right)  - \frac{E_q}{T} \right\}\,, 
\end{eqnarray} 
where the last term is the divergent vacuum contribution. As will be shown
below, this term influences the shape of the isentropic
trajectories near the phase boundary.

\begin{figure}[thbp]
\includegraphics[height=82mm]{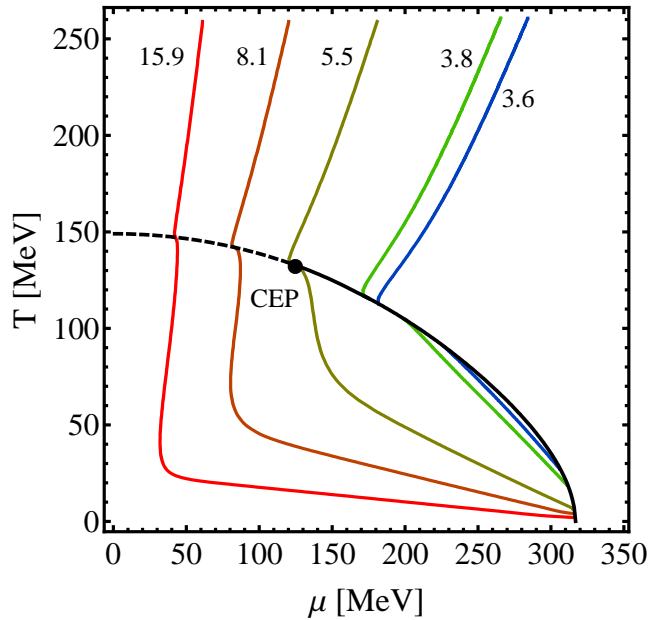}
\caption{Isentropes computed in the mean-field approximation to the
  quark-meson model, neglecting the vacuum term in \Eq{quark}. The
  values of $s/n$ are indicated at each isentropic trajectory. The
  chiral phase boundary, composed of a crossover and a first-order
  transition, is indicated by a broken and a full line, respectively.
  The bullet on the phase boundary indicates the position of the CEP
  in the MF approximation to this model.}
  \label{fig:isentropes_MF} 
\end{figure} 
 
In the MF approximation the expectation value $\vev \sigma$ is
determined by the corresponding classical equation of motion, the gap equation.
This is
obtained by minimizing the thermodynamic potential in the $\sigma$-direction
\begin{equation} 
  \left.\frac{\partial \Omega}{\partial \sigma}\right|_{\text{min}} =0 \ . 
\end{equation} 
The solution of the gap equation determines the $T$- and
$\mu$--dependence of the chiral order parameter $\vev \sigma (T,\mu)$
and the constituent quark mass $M_q=g \langle \sigma
\rangle$. The
expectation values of the pion fields $\vev{\vec \pi}$ vanish. We
have chosen the model parameters in the MF analysis so as to reproduce
the same vacuum physics as in the FRG approach.


\begin{figure}[thbp]
\includegraphics[height=82mm]{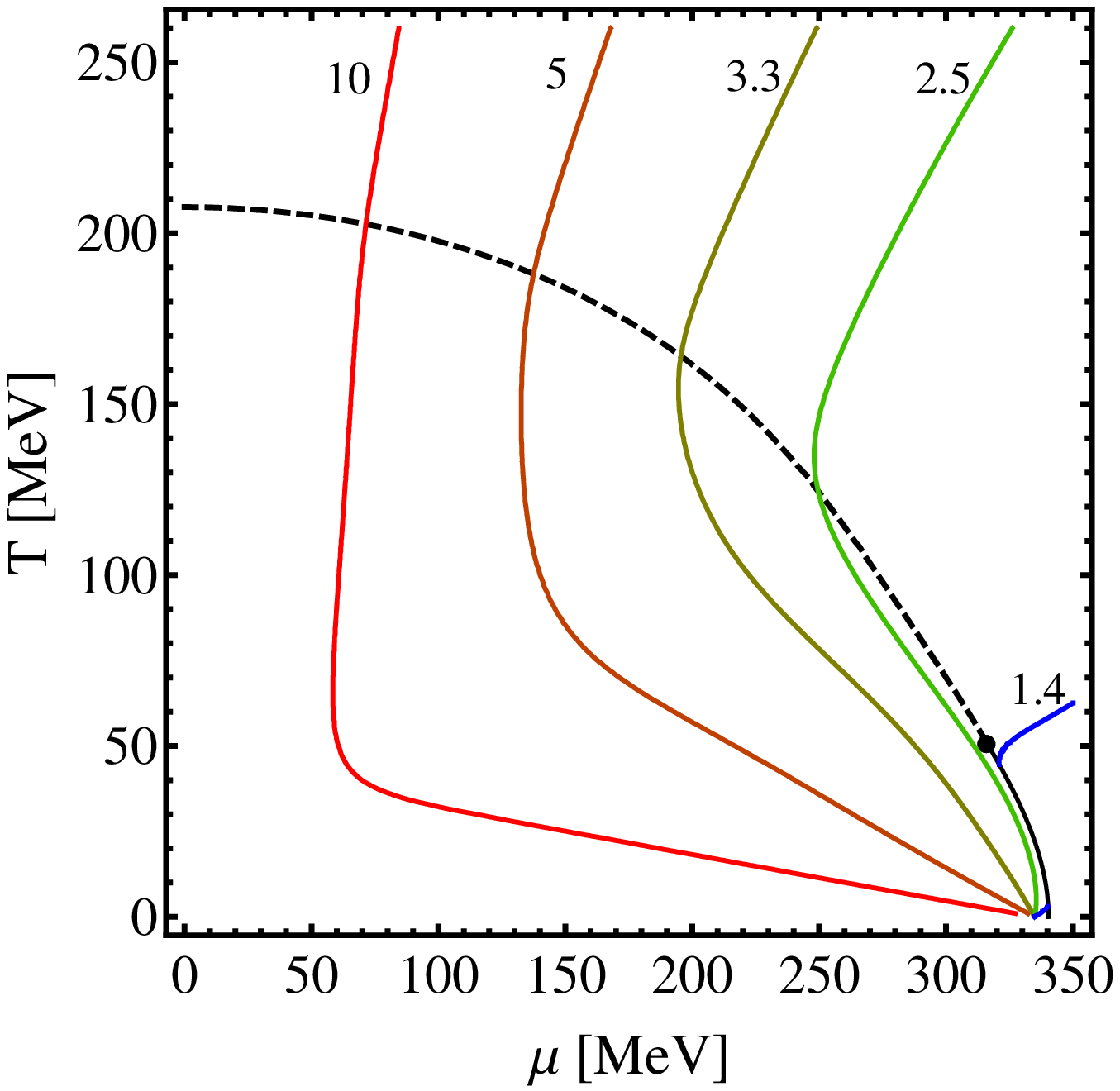}
\caption{The same as \Fig{fig:isentropes_MF}, but including the vacuum
  term in \Eq{quark}.}
  \label{fig:isentropes_MFD} 
\end{figure} 

In Fig.~\ref{fig:isentropes_MF} and \ref{fig:isentropes_MFD} we show
the isentropes obtained in the mean-field approximation with and
without the vacuum contribution. The divergence of the vacuum term is
regularized by a momentum cutoff $\Lambda = 583$ MeV. The parameters of the
model were in both cases chosen so as to reproduce the vacuum physics discussed
in section\ref{sec:frg}. We note that the position of the CEP is shifted to
smaller temperature and larger values of the chemical potential,
namely from $(T_c,\mu_c)=(130,147)$ to $(51,316)$ MeV, when the vacuum
contribution is included. Furthermore, the crossover temperature at
$\mu=0$ is increased from approximately 150 to almost 210 MeV, while
the critical chemical potential at $T=0$ is changed by less than 10
\%.

Our results also show that the kink in the isentropic trajectories
found at the phase boundary in the calculation, where the vacuum
contribution is neglected, is to a large extent an artifact of the
approximation. The thermal part of (\ref{quark}) yields a term of the form
$M_q^4 \log M_q^2$. With such a
term in the effective potential, the phase transition in the chiral
limit is first order for all densities, from $\mu=0$ and
$T\approx 200$ MeV to $T=0$ and $\mu\approx 300$ MeV~\cite{BBF}, as
recently found in \Ref{Schaefer:2006ds}. Due to the non-zero latent
heat, the isentropic trajectories are discontinuous at the phase
boundary in the $T-\mu$ plane. When a sufficiently strong explicit
symmetry breaking term is introduced, the first order transition is
smoothened into a crossover and the discontinuity in the isentropes
shows up as kinks at the phase boundary. However, when the vacuum
  term is included, the logarithmic term is cancelled and the
transition in the chiral limit is second order in flavor $SU(2)$.
Consequently, in the chiral limit the isentropes are continuous and
when the explicit symmetry breaking term is turned on the isentropic
trajectories remain smooth, as seen in \Fig{fig:isentropes_MFD}.

These considerations offer a plausible explanation for the different
behavior of the isentropes near the phase boundary in the flavor
$SU(2)$ versions of the quark-meson and Nambu--Jona-Lasinio (NJL))
models found in \Ref{Scavenius:2000qd}. In the quark-meson model the
vacuum term was dropped, while in the NJL model it must be included,
since the model otherwise does not yield a state with spontaneously
broken chiral symmetry. Consequently, one expects a kink at the
transition in the quark-meson model and smooth isentropes in the NJL
model, in agreement with the results of \Ref{Scavenius:2000qd}. 
We conclude that the strong kink structure in the crossover region,
which is obtained by neglecting the quark vacuum loop, is
unphysical.

In two recent papers, the isentropes have been computed in effective
chiral models including the coupling to the Polyakov loop in the
mean-field approximation~\cite{Kahara:2008yg,Fukushima:2009dx}. Also
in this case, the singular term cancels between the thermal and vacuum
contributions. Hence, one expects a kink in the quark-meson model with
Polyakov loop and smooth trajectories in the Polyakov--NJL (PNJL)
model. This is confirmed on a qualitative level by the calculation of
\Ref{Kahara:2008yg} and similar trajectories have been obtained in the
flavor $SU(3)$ version of the PNJL model in \Ref{Fukushima:2009dx}.
The cause of a less pronounced structure found in the isentropes
obtained in the PNJL model as the CEP is approached, remains
unclear.
A possible connection with the first order transition in the chiral
limit of the flavor $SU(3)$ model \cite{Gavin:1994p220} is uncertain,
since similar trajectories are obtained in the two flavor PNJL model,
where in the same limit the transition is second
order~\cite{Kahara:2008yg}.

At the first-order transition to the broken symmetry phase, the
isentropes are deflected to larger $\mu$ and smaller $T$, away from
the CEP, because the entropy per baryon in the co-existence region is,
in chiral effective models lacking gluonic degrees of freedom, smaller
in the symmetric phase. This may be different if the chiral and
deconfinement transitions coincide at finite $\mu$, since in QCD the
gluons, are expected to contribute a major part of the entropy per
baryon. However, if the so called quarkyonic
phase~\cite{pisarski-mclerran} is realized, the isentropes at the
first-order transition from hadronic to quarkyonic matter may be
similar to those found in the quark-meson model.

 
\section{Isentropic trajectories in the RG approach} 
\label{sec:isentropic}

As indicated in the introduction, the influence of fluctuations on the
isentropic trajectories at the chiral transition and in particular at
the critical endpoint of QCD is of high current
interest~\cite{Nonaka:2004pg, Asakawa:2008ti, Luo:2009sx}. We address
this problem by comparing the isentropic trajectories obtained using
the FRG approach with those obtained in the MF approximation in the
preceding section.

\begin{figure}[t]
    \includegraphics[height=82mm]{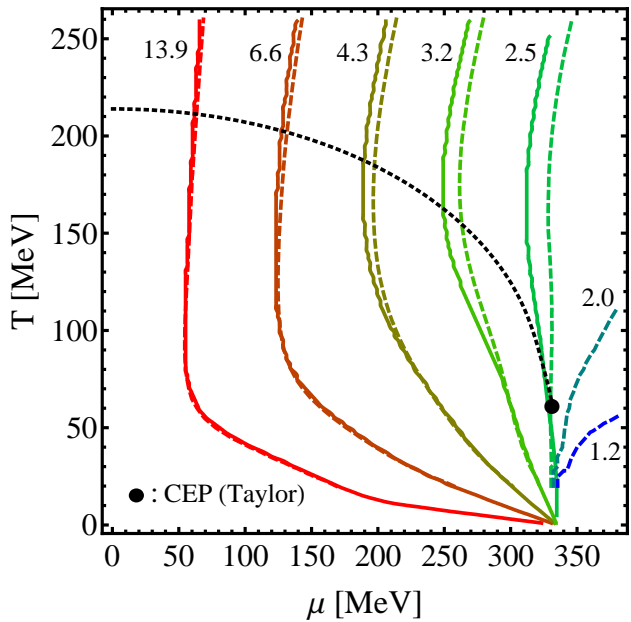}
\caption{Isentropes calculated in the quark--meson model using the \frgt$\ $and
\frgg$\ $methods (see text). The isentropes computed within the \frgt$\ $
approach are shown as solid lines whereas those
obtained within the grid method \frgg$\ $are indicated by dashed lines. The
$s/n$-ratio of each isentrope is indicated at each contour. The phase boundary,
with the CEP, obtained in the FRG approach, is indicated as in
Fig.~\protect\ref{fig:isentropes_MF}. The CEP shown is that obtained using the
Taylor
expansion method \frgt$\ $.}
\label{fig:isentropes_RG}
\end{figure}

Using the thermodynamic potentials obtained in the FRG approach, we
determine the phase diagram of the model \cite{Schaefer:2004en}. The
crossover line corresponds to a maximum of the chiral susceptibility.
In \Fig{fig:isentropes_RG} we show the resulting phase diagram
together with the corresponding isentropes. As in the MF approach, the
phase diagram has a generic structure and exhibits a critical
endpoint, which separates the first-order phase transition from the
rapid crossover. At the CEP, the transition is of second-order and
belongs to the universality class of the Ising model in three
dimensions.

A comparison with the MF results shows that the FRG phase diagram
is very similar to that obtained in the mean-field calculation with
the vacuum term included. In particular, the position of the critical
endpoint, the crossover temperature at $\mu=0$ and the critical
chemical potential at $T=0$ are almost identical. Thus, for the
quark-meson model, the effect of fluctuations on the phase diagram are
fairly small, once the fermion vacuum term is taken into account. With
our choice of model parameters the FRG approach yields approximately
$(T_c,\mu_c)=(14,328)$ MeV within the grid method (\frgg), while using
the Taylor expansion method (\frgt) we find $(T_c,\mu_c)=(64 ,336)$
MeV, as shown in \Fig{fig:isentropes_RG}. Thus, the location of the
CEP is not only model dependent~\cite{stephanov:2004wx}, but also
depends on the way the dynamics is implemented in the renormalization
group flow equation. The two approaches differ in the treatment of
higher order, irrelevant operators. This does not affect universal
quantities, like critical exponents, but it can affect non-universal
quantities, like the location of the critical endpoint.

A shift of the CEP to larger values of the chemical potential and
smaller temperatures was also found in the proper-time renormalization
group approach relative to the mean-field solution in
\Ref{Schaefer:2006ds}. In view of our results, it is likely that a
major part of this shift is due to the fermion vacuum term, which was
not included in the mean-field approximation.

We also show the isentropic trajectories in the ($T , \mu$)-plane,
obtained in the \frgt~ and in the \frgg~ approaches. The isentropes
show a smooth behavior everywhere, and agree qualitatively with the
mean-field trajectories shown in \Fig{fig:isentropes_MFD}, i.e., those
obtained with the vacuum term. It is reassuring that the \frgt~ and
the \frgg~ methods lead to a very similar isentropic trajectories.
Some deviations at large $\mu$, seen in Fig.~\ref{fig:isentropes_RG},
indicate limitations of the Taylor expansion method due to the
truncation of the series (\ref{eq:taylor}) at third order. Recent LGT
results \cite{lgtiso} on the isentropic trajectories near the QCD
crossover line show qualitatively a very similar smooth behavior as
obtained in our model with the FRG approach and shown in
Fig.~\ref{fig:isentropes_RG} and \ref{fig:isentropes_MFD}.

\begin{figure}[thbp]
\includegraphics[height=82mm]{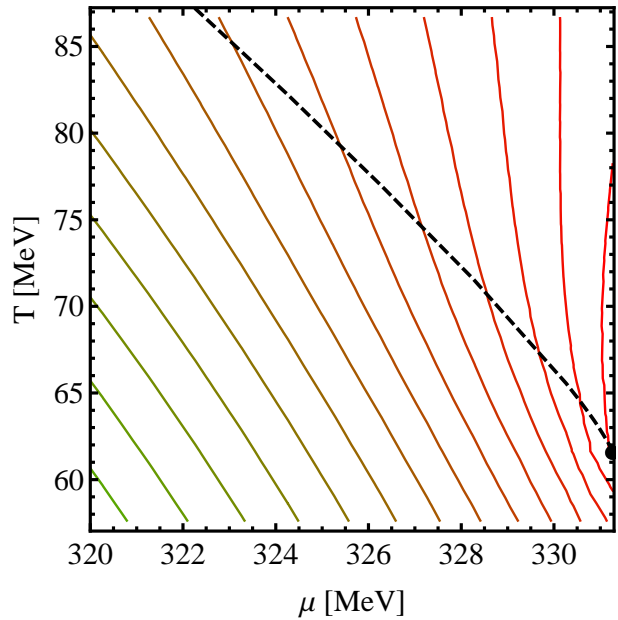}
\caption{Isentropes as in Fig.~\protect\ref{fig:isentropes_RG} but
  in a narrower region around the CEP from the side of the crossover
  transition (dashed line). The CEP is indicated as a black point at the right
edge of the figure.}
  \label{fig:BlowupCEP} 
\end{figure}

 
A comparison of the FRG and MF results shows no qualitative change of
the isentropic trajectories in the vicinity of the CEP. In particular,
the isentropes remain smooth also when the effect of long-wavelength
fluctuations is consistently included and show no sign of focusing
towards the CEP. This is clearly illustrated in
Fig.~\ref{fig:BlowupCEP}, which shows a blow-up of the FRG isentropes
in the vicinity of the CEP. Since the quark-meson model is in the same
universality class as QCD, our results do not support the
universality of the focusing phenomenon conjectured in
Refs.~\cite{Nonaka:2004pg, Asakawa:2008ti}. We stress that the RG
treatment of fluctuations employed here, reproduces the $Z(2)$
universal scaling of the relevant physical observables at the CEP
\cite{Schaefer:2006ds}.
 
The fact that the focusing effect is not universal can be understood
in general terms. The point is that the entropy and the baryon density
are both obtained as first derivatives of the thermodynamic potential
$\Omega$, which remain finite at the CEP, since only second- and
higher-order derivatives diverge. Consequently, the singular part of
the entropy per baryon does not diverge at the CEP and hence is not
guaranteed to dominate over the regular background contribution. It
follows that the isentropic trajectories are not universal, since they
depend on the relative strength of the universal singular part and the
non-universal background. In other words, the characteristic shape of
the isentropes in the vicinity of the CEP can vary from model to
model, even though they belong to the same universality class. The
model constructed in Refs.~\cite{Nonaka:2004pg, Asakawa:2008ti} yields
focusing of the isentropes towards the CEP because the singular part
of the thermodynamic potential is chosen by hand to be very large. In
chiral models, where the critical region around the CEP and around the
$O(4)$ transition line is small \cite{Schaefer:2006ds, SFR}, and
consequently the relative strength of the singular part of $\Omega$ is
small, it is unlikely that the focusing effect of the isentropic
trajectories reported in \cite{Nonaka:2004pg} can be observed.

For completeness we note that the fact that this effect is not
universal does not exclude the possibility that in QCD the relative
strength of the singular part is large. If this is the case, focusing
may then potentially be relevant for nucleus-nucleus collision
experiments~\cite{Asakawa:2008ti, Luo:2009sx}. However, then a
detailed study of the equilibration of long-range fluctuations in an
expanding system, along the lines of Ref.~\cite{Berdnikov:1999ph} is
needed in order to decide whether the isentropic trajectories are
relevant or not, as discussed in the introduction.

\section{Summary} 
\label{sec:summary}

The isentropic trajectories (contours of fixed entropy per baryon) in
the QCD phase diagram describe possible paths of the hydrodynamic
evolution of a thermal medium created in nucleus-nucleus collisions.
We investigated the behavior of the isentropic trajectories within the
chiral quark-meson model for two-quark flavors. The thermodynamics was
formulated using functional renormalization group (FRG) techniques and
the results were compared with two variants of the mean-field (MF)
approximation, one neglecting and the other one including the
fermion vacuum term.

Our studies of the isentropic trajectories near the chiral phase
transition were motivated by recent findings that the chiral critical
endpoint (CEP) acts as an attractor for the isentropes, leading
to a focusing towards the CEP \cite{Nonaka:2004pg}. It was argued
that the focusing effect would have important phenomenological
consequences for nucleus-nucleus collisions \cite{Asakawa:2008ti,
  Luo:2009sx}.
 
A comparison of the MF and the FRG results for the isentropes show
that the kink structure in the transition region, observed in some MF
calculations, is washed out when the fermion vacuum
contribution is properly included. Furthermore, in spite of the fact
that the entropy density and the baryon-number density rise rapidly
near the crossover transition and also as the CEP is approached,
the isentropes around the CEP remain very smooth. These
results raise doubts concerning the focusing of isentropes and its
phenomenological consequences.

Although, as already mentioned,
the isentropic trajectories of our model differ from those of QCD, 
the singular part of the thermodynamic potential near the CEP is universal. 
Our results show that in the quark-meson model the singular contribution to 
the entropy per baryon near the CEP is subdominant and hence that the
conjectured focusing
of the isentropic trajectories towards the CEP is not universal. However, 
the possibility still remains that in some systems of the same universality 
class focusing may appear, 
if accidentally close to the CEP the singular term dominates. 
Thus, we conclude that the existence of such an effect in QCD matter is unlikely 
but not completely excluded.

\subsection*{Acknowledgments} 
We acknowledge stimulating discussions with Jochen Wambach and Vladimir Skokov. 
KR received partial support from the Polish Ministry of Science and the Deutsche
Forschungsgemeinschaft (DFG) under the Mercator Programme.

\end{document}